\documentclass[aps,pra,english,twocolumn,showpacs]{revtex4-2}

\def\be{\begin{equation}}
	\def\ee{\end{equation}}
\def\bea{\begin{eqnarray}}          
	\def\eea{\end{eqnarray}}
\def\bi{\begin{itemize}}
	\def\ei{\end{itemize}}

\usepackage{xcolor}
\usepackage{graphicx}

\usepackage{float, ragged2e, booktabs}
\usepackage[colorlinks=true, citecolor=blue, urlcolor=blue ]{hyperref}
\usepackage{amsmath,amstext,amssymb,babel,bm,color,times}

\usepackage{physics}
\usepackage{tikz, adjustbox}
\usetikzlibrary{quantikz}
\tikzcdset{scale cd/.style={every label/.append style={scale=#1},
		cells={nodes={scale=#1}}}}

\usepackage{subcaption}

\usepackage{etoolbox,tcolorbox}

\apptocmd{\thebibliography}{\justifying}{}{}
\makeatletter
\long\def\@makecaption#1#2{%
	\vskip\abovecaptionskip
	\sbox\@tempboxa{#1. #2}%
	\ifdim \wd\@tempboxa >\hsize
	\begin{justify}
		#1. #2\par
	\end{justify}
	\else
	\global \@minipagefalse
	\hb@xt@\hsize{\hfil\box\@tempboxa\hfil}%
	\fi
	\vskip\belowcaptionskip}
\makeatother

\begin{document}
	
	
	\title{Restricted Randomized Benchmarking with Universal Gates of Fixed Sequence Length}
	
	
	\author{Mohsen Mehrani$^{1}$, Kasra Masoudi$^{2}$, Rawad Mezher$^{3}$, Elham Kashefi$^{4,5}$, and Debasis Sadhukhan$^{5}$}
	
	
	\affiliation{\(^1\) Institute for Quantum Science and Technology, University of Calgary, 2500 University Drive NW, Calgary, Alberta T2N 1N4, Canada}
	\affiliation{\(^2\) Department of Mathematics, Simon Fraser University, 8888 University Drive, Burnaby, BC, V5A 1S6, Canada}
	\affiliation{\(^3\) Quandela SAS, 7 Rue L\'{e}onard de Vinci, Massy 91300, France}
	\affiliation{\(^4\) Laboratoire d’Informatique de Paris 6, CNRS, Sorbonne Universit\'{e}, 4 Place Jussieu, Paris 75005, France}
	\affiliation{\(^5\) School of Informatics, University of Edinburgh, 10 Crichton Street, Edinburgh EH8 9AB, United Kingdom}


	\begin{abstract}
		The standard randomized benchmarking protocol requires access to often complex operations that are not always directly accessible. Compiler optimization does not always ensure equal sequence length of the directly accessible universal gates for each random operation. We introduce a version of the RB protocol that creates Haar-randomness using a directly accessible universal gate set of equal sequence length rather than relying upon a \textit{t-design} or even an approximate one. This makes our protocol highly resource efficient and practical for small qubit numbers. We exemplify our protocol for creating Haar-randomness in the case of single and two qubits. Benchmarking our result with the standard RB protocol, allows us to calculate the overestimation of the average gate fidelity as compared to the standard technique. We augment our findings with a noise analysis which demonstrates that our method could be an effective tool for building accurate models of experimental noise.  
			\end{abstract}
	
	\maketitle

	\section{Introduction}
	\label{sec:intro}
	
	Scalable quantum hardware has witnessed formidable progress over the last decade led by efforts from academia and major industries \cite{Brooks2023}. While quantum processors today are large enough to contain a few hundred qubits, their usefulness to showcase quantum advantage \cite{https://doi.org/10.48550/arxiv.1203.5813} is severely challenged by its sensitiveness to environmental noise and imperfect operations \cite{Preskill2018quantumcomputingin, https://doi.org/10.48550/arxiv.2203.17181}. The recent advancement in quantum technology made quantum computers accessible via the internet-cloud which allows users to use quantum computers in a hardware-agnostic way. However, since current hardware suffers from a diverse array of errors,  benchmarking these devices is crucial to understand the efficiency of quantum processors.
	
	While in the early stages of quantum processors the total number of available controllable qubits used to be the advertised number to showcase the strength of a quantum device, it was soon realized that the noise in the device should also be incorporated within the reported single number metric. Hardware connectivity should also be an important factor because in hardware with limited connectivity, one would need extra swap gates to implement two-qubit gates on distant qubits. In pursuit of such a single metric, concepts like {\it quantum volume} \cite{PhysRevA.100.032328, BlumeKohout2020volumetricframework, Jurcevic_2021} emerged where together with the total size of the circuit, the depth of the circuits, that can be executed with noise resilience, was also considered. Such metric however cannot ensure error resilience of a particular application. This inspired various works \cite{Lubinski2021-nh, 2403.12205} towards better application-oriented performance benchmarks and metrics such as the {\it algorithmic qubit} \cite{Maksymov2023-jp} that uses an algorithm-agnostic approach. 
	
	While these kinds of summary metrics are appealing to non-specialists for portraying the holistic performance of a quantum machine, they fail to capture the full complexity of errors in the quantum hardware. This is where component-level specifications come in. Quantum computers suffer from a broad spectrum of errors, such as state preparation and measurement (SPAM) errors, gate operation errors, cross-talks, errors due to loss of energy (T1 time) or phase (T2 time) coherence, etc. Benchmarking errors at such a low level is important to specialists because it allows them to isolate their sources which in turn can be comprehended to enhance the device's performance. 
	Moreover, such low-level benchmarks must be hardware-agnostic, otherwise understanding how they interfere and contribute to the single summary metric will become problematic \cite{Proctor2021, PhysRevA.99.032318}.    
	Although errors corresponding to single- or two-qubits can be analyzed in detail using tomographic techniques \cite{Nielsen2021gatesettomography, PRXQuantum.2.040338}, multiqubit circuits often face more intricate issues, such as crosstalk \cite{PhysRevLett.109.240504,PhysRevLett.123.030503,Sarovar2020detectingcrosstalk,Maciejewski2021modelingmitigation,Proctor2021,https://doi.org/10.48550/arxiv.2003.02354}, for which these approaches are either inadequate or become prohibitively costly due to exponential increases in sampling complexity. 
	
	A well-established method for assessing the average quality of operations in quantum hardware that is independent of SPAM errors is {\it randomized benchmarking } (RB) \cite{Emerson_2005}. It provides a reliable estimate of the magnitude of the average error of a random operation robustly in any digital quantum device. In its original manifestation \cite{Emerson_2005}, one is required to access all possible Haar-random unitary operations that need to be executed over multiple sequence lengths to amplify small errors in the {\it gate set}. The digital quantum computers that we have today are all universal in the sense that all possible operations are accessible via some standard universal gate set given that we are allowed to apply the standard gates as many times as required. 
	The Solovay-Kitaev theorem \cite{Kitaev1997, Kitaev2002, https://doi.org/10.48550/arxiv.quant-ph/0505030},
	when generalized for $n$ qubits, provides an estimate: an arbitrary $n$ qubit quantum operation $\mathcal{U}(2^n)$, when decomposed into single- and two-qubit gates of an universal gateset, would need an exponentially large ${ O}(n^22^{2n}\log(n^22^{2n}/\epsilon))$ number of such gates to mimic $\mathcal{U}(2^n)$ upto an accuracy $\epsilon$. Such exponentially long gate sequences for each Haar-random unitaries within a RB protocol poses a significant challenge to the near-term quantum hardware as they can only execute shallow-depth circuits with noise resilience.  
	
	However, people soon came up with RB protocols using finite groups \cite{PhysRevA.77.012307, PhysRevA.80.012304, PhysRevA.90.030303, CarignanDugas2015CharacterizingUG, Cross2016, PhysRevLett.123.060501, https://doi.org/10.48550/arxiv.1806.02048, Erhard2019} such as {\it Clifford groups} \cite{PhysRevA.77.012307, PhysRevLett.106.180504, PhysRevA.85.042311} that can mimic Haar-random operations. RB based on unitary 2-design generates gate sequences of different lengths, and at the end, another inverse circuit is applied to nullify all previous operations. By measuring in the computational basis, the fidelity between the measured output and the ideal outcome can be computed. This procedure is repeated for gate sequences of multiple lengths. By plotting the fidelity against the sequence length, one can extract information about the average error rate of the quantum device. 
	
\begin{figure}[t]
		\includegraphics[width = 0.95\linewidth]{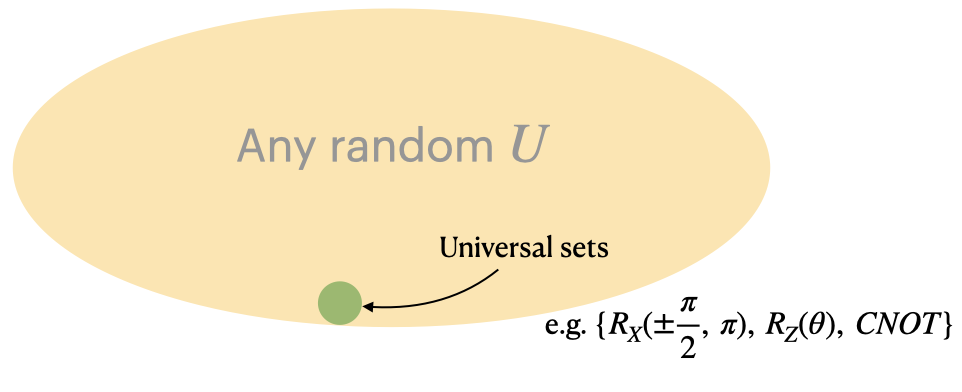}\\
		\includegraphics[width = 0.95\linewidth]{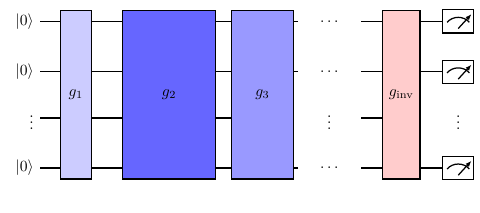}
		\caption{(upper panel) Randomized benchmarking protocol requires one to implement any Haar-random operations $U$ from the $n$-qubit operator space $\mathcal{SU}(2^n)$. In any digital quantum computer, such operations are accessible via the universal standard single and two-qubit gates set by the manufacturer. A quantum device comes with a compiler that decomposes the $n$-qubit operation $U$ into a sequence of single and two-qubit gates that are natively accessible in the device. The compiler optimization, however, does not ensure equal sequence length for all the operations. (lower panel) For a typical RB step, it implements a Haar-random operations using gateset $g_1$ of shallow depth whereas for the next, it uses a much deeper circuit of gateset $g_2$.  }
		\label{fig:schematic}
	\end{figure}

	In the standard RB, the Haar-random operations should be directly accessible. However, this is not quite practical. The current quantum devices rely on a compiler to obtain universal gate decompositions which are directly accessible to the device. For typical Haar-random operations $\{U\}$ , the compiler optimization does not ensure equal sequence lengths of the universal gate decompositions. While for a typical $U_1$, the compiler could find a shallow circuit $g_1$, it may have to use a comparatively deep circuit $g_2$ for another operation $U_2$ (see Fig.\ \ref{fig:schematic}). Such disparities often result in an overestimation of the average error rate of the device if one sticks to the standard RB technique. In this work, we devised a new version of the protocol, which we call {\it Restricted Randomized Benchmarking} where for each Haar-random operation we use universal gate decomposition of equal sequence length. 
	
	In this work, we try to bridge the gap between the theory of RB and its experimental realization. We explicitly illustrate our results in the case of single- and two-qubit scenarios. Assuming the directly accessible standard universal gateset to be $\{ R_X(\pm \frac{\pi}{2}, \pm \pi), R_Z(\theta), $C-NOT$\}$ which is inspired from the native gate-sets of the superconducting hardware, we set out a protocol to generate Haar-random unitaries using direcly accessible gates of fixed sequence length.  For single qubits, any random operations can be expressed using 4 sequences of universal gates while for the two-qubits operations, results from Ref.\ \cite{PhysRevA.69.032315, PhysRevA.69.010301} guarantee that only three C-NOTs together with a few single qubit rotations are enough. Using these results, we construct our single- and two-qubit RB gate sequences of equal length for all applied Haar-random operations. This in turn allows us to calculate the overestimation of the average gate fidelity as compared to the standard techniques. We also benchmark our results with the RB protocols based on 2-design which shows a close match. 
	For small qubit numbers, where the resources required to generate Haar randomness are still manageable, we argue that the protocol introduced here will lead to a resource-friendly and compiler-aware technique for performing RB on existing quantum hardware. 
	
	The results obtained allow us to examine the noise characteristics inherent in the experimental hardware. The standard RB, in its original zeroth-order version, assumes a gate-independent noise in the hardware to establish an exponential decay of the average fidelity. However, in practice, the experimental hardware might contain gate-dependent noise, and therefore the exponential decay is not guaranteed.  
	We supplement our findings through a noise analysis of our restricted RB method, demostrating with a typical example that the exponential decay we observe could be accounted for if the prevailing noise in the hardware is depolarizing. Although this explanation may not be exclusive, we believe our method could be an effective tool for building efficient and quantitatively accurate models of experimental noise in quantum hardware. An improved noise model is anticipated not only to pave the way for efficient error detection and mitigation strategies but also to aid in achieving the long-term goal of quantum error correction in quantum computing. 

	\section{Randomised Benchmarking: Preliminaries}
	
	An RB protocol typically requires three steps. In the first step, the quantum register is initialized with a simple quantum state such as $\rho = |\psi\rangle\langle\psi|$.  In the second step, a sequence of random unitary quantum operations $\{U_i\}_{i=1,2,\ldots,m}$ of varying length $m$ is applied which is then finished by an inverse operation $U_{m+1}^\text{inv}$ to nullify all previous random operations. 
	Because of the presence of the inverse operation at the end, the entire chain of operations of length $m+1$ turns out to be identity in the noiseless scenario and therefore we should measure the initial state with unit fidelity. 
	In the presence of noise, however, the applied gate sequences are not exact. The noise acting on the gates corrupts their operations which in turn 
	reduce the final fidelity to something less than one. 
	We repeat this procedure for different values of $m$ and record the fidelity.
	In the final step, we extract the average error rate by fitting the recorded data against sequence length $m$ which represents the error rate of implementing gate operations and is free from other sources of errors such as SPAM errors that do not depend on the length of the sequence. 
	
	\subsection{Haar-Twirl}
	\label{sec:haartwirl}
	
	Let us first consider the scenario for $m=1$. Suppose $U$ represents any Haar-random quantum operation within the n-qubit unitary space $U\in \mathcal{U}(2^n)$, with ${\hat{\bf \Lambda}_U} \equiv {\Lambda}_U \circ {U}$ describing the noisy experimental realization of $U$. 
	In general, ${\Lambda}_U$ might represent any completely positive and trace-preserving (CPTP) superoperator that characterizes the noise associated with the experimental execution of the gate operation $U$ and can be described by ${\Lambda}_U (\rho) = \sum_k A_k \rho A_k^\dagger$ where $A_k$ are the Kraus operators. 
	We aim to determine how different the actual experimental implementation $\hat{\bf \Lambda}_U$ is from the ideal operation $U$. To calculate the same, one can look at the average gate fidelity,
	\bea
	F(U, \hat{\bf \Lambda}_U) = \int d\rho F(U \rho U^\dagger, \hat{\bf \Lambda}_U(\rho)),
	\eea 
	where the fidelity is given by $F(\sigma,\tau) = \Tr[\sqrt{\sigma^{1/2}\tau\sigma^{1/2}}]^2$, $\sigma$ and $\tau$ are density matrices. For pure states, this reduces to $F({U}, \hat{\bf \Lambda}_U) = \int d\psi \langle \psi| U^\dagger \hat{\bf \Lambda}_U(|\psi\rangle\langle\psi|) U|\psi\rangle$ where the average is over all pure states. 
	
	The average fidelity of hardware is then calculated by averaging over all the unitaries, given by 
	\bea
	\bar{F} = \int_{U \in \mathcal{U}(2^n)}d\mu(U) F({U}, \hat{\bf \Lambda}_U) 
	\label{eq:avFidelity}
	\eea
	where the integral is over the Haar measure $d\mu$ \cite{Pozniak1998} on unitary space $\mathcal{U}(2^n)$. The average error rate of the device is simply $1-\bar{F}$. 
	
	One can now think of a motion reversal transformation to see the equivalence $F({U},\hat{\bf \Lambda}_U)=F(\mathbb{I},{ \Lambda}_{U^\dagger U})$ where ${ \Lambda}_{U^\dagger U}$ is the noise channel associated with the motion reversal experiment $\hat{\bf \Lambda}_{U U^\dagger}\equiv {\Lambda}_{U U^\dagger}\circ {UU^\dagger}$. Nielsen \cite{NIELSEN2002249} pointed out that fidelity of a superoperator $\Lambda$ with identity is same as the fidelity of its Haar twirl with identity $F(\int_{U}d\mu(U) U^\dagger \Lambda U, \mathbb{I}) = F({\Lambda},\mathbb{I})$ which lead us to  
	\bea
	F({U},\hat{\bf \Lambda}_U)=F(\mathbb{I},{\Lambda}_{U^\dagger U})=F(\mathbb{I},{\Lambda}_{av})
	\eea
	where ${\Lambda}_{av} = \int_{U'}d\mu(U') {U'}^\dagger \circ {\Lambda}_{U^\dagger U} \circ {U'}$ is the Haar-twirl of the superoperator ${\Lambda}_{U^\dagger U}$. Emerson {\it et al.} \cite{emerson2005scalable} came up with this formalism and showed that ${\Lambda}_{av}$ is unitary invariant i.e.\ ${\Lambda}_{av}={U} \circ {\Lambda}_{av} \circ {U}^\dagger ~ ~ \forall~~ U \in \mathcal{U}(2^n)$ and have two irreducible representations. Applying Schur's lemma on each of them gives the form of the Haar-twirled superoperator 
	\bea
	{\Lambda}_{av} (\rho) = p\rho + (1-p) \Tr(\rho) \frac{\mathbb{I}}{2^n}  = {\Lambda}_d(\rho)
	\label{lambda_d}
	\eea
	with $p = \left(\Tr({\Lambda}_{av})-1\right)/(2^{2n}-1)$ which is just a depolarizing channel.  
	
	Therefore, for a single-layer randomizing benchmarking protocol, the average fidelity of a device (for all possible operations) depends on a single parameter. However, when we extend this protocol for a sequence of length $m$, we must make some assumptions about the noise channel to obtain the single parameter characterizable form. 
	In general, for a m-sequence RB, the average fidelity is given by,
	\bea
	&&\bar{F} =\int d\mu(U_1) \ldots d\mu(U_m) \nonumber \\
	&&\langle \psi | \hat{\bf \Lambda}_m \circ  \hat{\bf \Lambda}_{U_{m+1}^\text{inv}} \circ \hat{\bf \Lambda}_{{U}_{m}}\circ\hat{\bf \Lambda}_{U_{m-1}}\cdots\hat{\bf \Lambda}_{{U}_1}(|\psi\rangle\langle\psi|)|\psi\rangle, \nonumber
	\eea
	where $\hat{\bf \Lambda}_m$ is the experimental implementation of the state preparation and final measurement. If we assume that the noise channels associated with random operations including the inverse operation is time and gate-independent and represented by ${\Lambda}$, the average fidelity for each length $m$ become
	\bea
	&&\bar{F} =\int d\mu(U_1) \ldots d\mu(U_m) \nonumber \\
	&&\langle \psi|{\Lambda}_s \circ {U}_{m+1}^\text{inv}\circ{\Lambda}\circ {U}_{m}\circ{\Lambda}\circ {U}_{m-1}\cdots{\Lambda}\circ {U}_1(|\psi\rangle\langle\psi|)|\psi\rangle \nonumber \\
	&&=\langle \psi|{\Lambda}_{s,t} ({\Lambda}_d)^m|\psi\rangle,
	\eea
	Note that the error channel associated with the inverse operation has also been absorbed within ${\Lambda}_s = {\Lambda}_{m}\circ {\Lambda}$ where ${\Lambda}_{m}$ is the superoperator for SPAM errors. After twirling, this channel becomes ${\Lambda}_{s,t}$. It manifests itself as a depolarizing channel, as the state measurement is carried out in the basis $\{|\psi\rangle,|\psi_\perp\rangle\}$, where off-diagonal elements do not contribute. ${\Lambda}_d$ is the twirled version of the channel ${\Lambda}$ which also has the form of a depolarizing channel characterized by the depolarizing parameter $p$. Finally, the average fidelity becomes
	\bea
	\bar{F} = p_s p^m + (1-p_s p^m)/d = A + Bp^m
	\label{AB}
	\eea
	where $p_s$ characterizes the twirled depolarizing channel ${\Lambda}_{s,t}$ and the coefficients are $A=1/d $ and $B=(1-{1}/{d})p_s.$ This final form of the average fidelity allows one to extract the parameter $p$ as we plot the average fidelity with the sequence length $m$. The Haar-twirl plays the pivotal role here to distill the complex quantum channels into a comprehensive quantifiable form.


	
	
	\subsection{\lowercase{{\it t-design}} based RB protocol}
	
	Since implementing a long sequence of Haar-random unitaries to produce the Haar-twirl is inefficient, we can bypass the same by mimicking the Haar-twirl by {\it unitary t-designs}. The goal of \emph{unitary t-design} is to suggest a small set that duplicates the properties of the distribution over total unitary space with Haar-measure for any polynomials of degree t \cite{PhysRevA.57.127}. It contains a small finite set of unitaries $\{C_k\}_{k=1,\ldots,K}$ with the property
	\bea
	\frac1K \sum_{k=1}^K P_{t}(C_k) = \int_{U \in \mathcal{U}(2^n)} d\mu(U) P_t(U),
	\eea
	for any polynomial $P_t$ of degree t.
	Therefore, sampling from the same is equivalent to sampling from the Haar random unitaries up to order $t$. 
	In most cases of our interest, a 2-design suffices. 
	The 2-design is a uniform distribution over \emph{Clifford group} which is generated by $\expval{H, S, CNOT}$ (see Appendix \ref{append:clifford}). So, instead of calculating the average of fidelity at Eq.\ (\ref{eq:avFidelity}), we can sum over the Clifford members \cite{PhysRevA.80.012304}.
	\begin{align}
		\bar{F}= \frac{1}{K}\sum^{K}_{k=1} F(C_k, \hat{\bf\Lambda}_{C_k})
	\end{align}
	This is an extremely efficient approach 
	because any element of the Clifford group can be generated using only $\mathcal{O}(n^2/\log(n))$ single- and two-qubit gates as compared to the exponential scaling for any arbitrary Haar-random gate \cite{Emerson2003PseudoRandomUO}.
	
	\section{Restricted Randomized Benchmarking Protocol}
	
	Most state-of-the-art quantum hardware relies on its compiler to find the native gate decomposition of the target unitaries. For example, the random unitary $U_i \in U (2^n)$ is executed by a gate decomposition of depth d, given by $g_i = \{X^1_i~X^2_i~\cdots X^d_i\}$ where each $X$ is the tensor product of single- and two-qubit native gates acting over n-qubits. When implementing this, the error channel associated with each native gate does not commute, and therefore we do not have a simple error channel of the assumed form, ${\Lambda}_{U_i} \circ {U}_i$, instead it has a more complicated form $$(\Lambda_{X^1_i}\circ X^1_i)(\Lambda_{X^2_i}\circ X^2_i)~\cdots (\Lambda_{X^d_i}\circ X^d_i)$$ for the native gate sequence $g_i$ implementing $U_i$.  As the gates and the superoperator do not commute in general, one cannot take all native gate operations $X$'s to the right to get the Haar-random unirary $U_i$. The noisy implementation of the Haar-random $U_i$ does not have the form $\Lambda_{U_i} \circ U_i$ in this case. Therefore, implementing the Haar twirl to get the unitary invariant superoperator is not straightforward with native gate implementation. 
	Moreover, the depth $d$ for any two Haar-random operations can be different because of the compiler optimization, and therefore the associated noise channel would contain additional superoperators even if we assume them to be gate and time-independent. For these reasons, it is difficult to obtain the Haar-twirled superoperator $\Lambda_{av}$ which is a unitary invariant that was the prerequisite behind the success of randomized benhcmarking. 
	%
	%
	%
	%
	The \emph{t-design} based techniques are not immune to these shortcomings either as for each of its design elements $C_k$ we need to rely on the native gate sequences for their experimental implementation. Although they are efficient in terms of sample complexity, the actual runs are not coming from \emph{t-designs} anymore, instead they are an approximation of the device's noisy native gates.
	%
	%
	%
	%
	%
	Therefore, although they should mimic the Haar twirl in practice, their native gate implementation, being a noisy mixure, does not become unitary invariant.
	
	In this work, we try to address this gap between the theory of RB and its experimental implementation by ensuring that each implementation of the Haar-random unitary is consists of equal depth sequences. This allows us to treat the associated noises in equal footing and leads to a better estimate of the average fidelity as compared to the standard RB. Notice, however, that the non-commutativity between the noise channels and native gate operations still possesses a challenge to the unitary invariance of the Haar-averaged super operator. In Appendix \ref{append:gate_dependent_channel}, we present a numerical analysis of our considered twirl considering a typical noise channel composed of both coherent and incoherent components and discuss the circumstances under which unitary invariance may hold.

	

	Let us first set out the protocol that implements Haar-random unitaries $U \in \mathcal{U}(2^n)$  for $n=1,2$, by using native gate sequences of fixed depth. 
	We denote the parametric native gates of the device as $\{X^j(\alpha_j)\}_{\vec{\alpha}}$, where $\vec{\alpha}$ is a tuple that depends on the number of free parameters for an n-qubit unitary. The number of elements in the tuple in that case is $2^n-1$. 
	By establishing a framework for fixed-length decomposition of any unitary, we set conditions on the selection of the parameters $\alpha_j$ to guarantee Haar randomness. Therefore, choosing $\{\alpha_j\}$ from certain marginal distributions, we generate the Haar-random unitaries to be used for the RB protocol
	%
	%
	\begin{align}
		 \Pi_j X^j{(\alpha_j)} =  U(\Vec{\alpha}) 
	\end{align}
	Needless to say, the ordering of the sequence is important, otherwise, the correct representation would not be generated. However, this ordering might be invariant to unitary matrix choice as there might exist a universal circuit that can span the entire unitary space through its free parameters. We show this explicitly for single- and two-qubit cases. 
	
	\subsection{Fixed depth Haar-random single-qubit unitary }
	
	The universal gateset of Rigetti QCS \footnote{Rigetti Documentation: pyQuil, \url{https://pyquil-docs.rigetti.com/en/v2.7.0/apidocs/gates.html}} are $R_Z(\phi)$ with any arbitrary angle, $R_X(\theta)$ with angles $\theta = \pm \pi, \pm \pi / 2$ and $CZ$ gate. Single-qubit unitaries has three free parameters. 
	For our construction, we decompose a single-qubit unitary $A(\phi, \theta, \omega) $ as
	\begin{align}
		A(\phi, \theta, \omega) 
		&= 
		\begin{bmatrix} e^{-i(\phi + \omega)/2}
			\cos(\theta/2) & -e^{i(\phi - \omega)/2} \sin(\theta/2)
			\\ e^{-i(\phi - \omega)/2} \sin(\theta/2) & e^{i(\phi +
				\omega)/2} \cos(\theta/2) \end{bmatrix} \label{eq:arbitrary_single_unitary} \nonumber\\
		&= R_Z(\omega) 
		\cdot R_X(-\pi/2) R_Z(\theta) R_X(\pi/2) \cdot R_Z(\phi)
	\end{align}
	%
	
	\begin{figure}[h]
		\includegraphics[width = 0.48\textwidth]{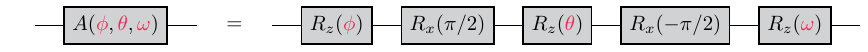}
		\caption{Native gate construction for arbitrary single-qubit unitary matrices with three free parameters.}
		\label{fig:universal_single_qubit_circuit}
	\end{figure}

	The key idea is to choose the joint probability distribution of the universal circuit parameters $\{\theta, \omega, \phi\}$ in such a way that the resulting unitaries become Haar random in the unitary space $\mathcal{U}(2)$.   
	%
	%
	%
	The Haar measure is rotation invariant. So, if one pure state is taken on the Bloch sphere and mapped by Haar-random matrices, the final state is expected to be equally likely any other point on the sphere surface. So, we expect the Haar measure to have a uniform surface density.
	
\begin{figure}[t]
		\includegraphics[width=0.9\linewidth]{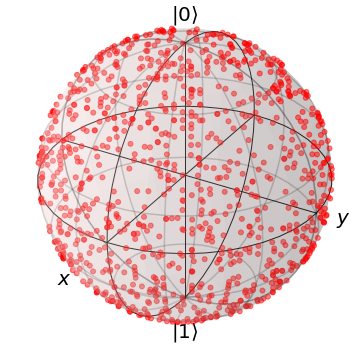}
		\caption{Uniformly distributed single-qubit states $A|0\rangle$ on the Bloch sphere. The uniformly distributed random states ensure the Haar randomness of the constructed single-qubit unitaries.}
		\label{fig:single-qubit-verification}
	\end{figure}

	Assuming a uniform density of the surface of the sphere, the correct distribution over $\{\phi, \theta, \omega\}$ can be found. $R_Z(\omega)$ and $R_Z(\phi)$ are simply a phase shift. As the distribution is symmetric to rotation along the z-axis, we expect a uniform distribution over these two angles. However, $R_{Y}(\theta) = R_X(-\pi/2) R_Z(\theta) R_X(\pi/2)$ is a beamsplitter. To sample $\theta$ uniformly at random in this context, we must sample from the distribution Prob($\theta$)=$\sin \theta$. By this intuition, we can easily deduce that the marginal distributions:
	\begin{align}
		d\mu_{H}(U) &= \sin\theta d\theta d\omega d\phi \\
		\rho(\phi,\omega) &= \frac{1}{2\pi} \\
		\rho(\theta) &= \frac{\sin(\theta)}{2} \label{eq:haar-random_dist_single_qubit}
	\end{align}
	So if the triplets $\{\omega, \theta, \phi\}$ are sampled by using these marginal distributions, it is guaranteed that the final circuit will be a Haar-random circuit. We verify the same by plotting the state $A|0\rangle$ on the Bloch sphere. Beginning with a constant initial state and subjecting it to a Haar random unitary, we anticipate that the resultant states $A|0\rangle$ will uniformly populate the Bloch sphere, which we confirm in Fig. \ref{fig:single-qubit-verification}. 
	
	Now that we have constructed a fixed-depth Haar-random single-qubit circuit for a single layer $m=1$, we can go for the RB protocol with different $m$. The fixed-depth circuits come into practice while generating random sequences of Haar random unitaries for the RB protocol. As we indicated earlier, Haar-random unitary transformations are implemented in a row plus the inverse of the total unitary transformation of the circuit. In the noiseless case, we are assured that the fidelity of the output state and the initial state would be maximally one. However, any quantum device encounters in-between noise which reduces the fidelity average.
	
\begin{figure}[h]
		\centering
		\includegraphics[width=0.99\linewidth]{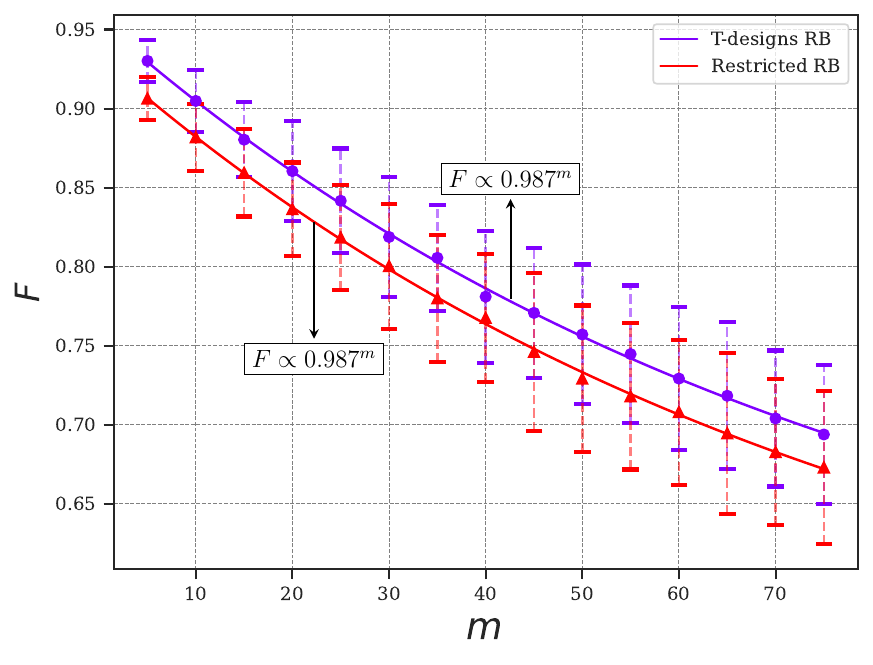}
		\caption{Comparative Analysis of RB schemes on single-qubit systems. - Here we compare two RB schemes - restricted RB and two-designs RB - applied on the Zeroth qubits of the Aspen-M-3 processor of Rigetti. Detailed experimental procedures are discussed in the results section. Error bars represent the standard deviation from 200 generated circuits with 800 shots each (refer to Table \ref{tab:rb_results} for detailed results). The purple curve illustrates the extrapolated exponential fit for the two-designs RB, given by the equation $F = 0.422 \times 0.987^m + 0.534$. Similarly, the red curve represents the fit for the restricted RB, described by $F = 0.425 \times 0.987^m + 0.508$. 
		}
		\label{fig:rb_single_qubit}
	\end{figure}

	In Fig.\ \ref{fig:rb_single_qubit}, the average fidelity (F) versus sequence length (m) can be observed. For simplicity, we use $F$ to denote average fidelity instead of $\bar{F}$ in the remainder of the paper. Motivated by Eq. (\ref{AB}), we fit the same to an exponential curve, which shows an excellent match. 
	Moreover, the obtained fidelity using restricted RB matches up to 3 decimal points to the obtained result using the {\it t-design} version. This is expected for single-qubit because the compiler optimization for the single-qubit unitaries is always of fixed depth. However, when comparing with the calibration data (see Table \ref{tab:rb_results}), we see that the reported data is a little overestimated than our findings.

	\subsection{Fixed depth Haar-random two-qubit unitary }
	
	
	
\begin{figure}[h]
		\includegraphics[width = 0.45\textwidth]{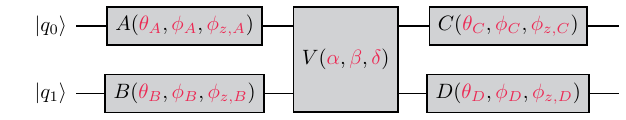}
		\caption{Construction of any two-qubit unitary $U$ with 15 control parameters. $A, B, C$ and $D$ are single-qubit gates while $V$ is entangling two-qubit gate with 3 free parameters. Refer Fig.\ \ref{fig:v_gate_circuit} for the native gate implementation of $V$.}
		\label{fig:universal_circuits}
	\end{figure}

	Let us now proceed towards the main result of our work. For any multi-quit random unitaries, the compiler optimization would produce native gate sequences of different depths that tries to minize the number of gate used. To the best of our knowledge, there exists no protocol that can guarantee a fixed depth (and composed of 1 and 2-qubit gates) circuit for generating any n-qubit random operation $U \in \mathcal{U}(2^n)$, although interesting progress in this direction has been made in Ref.\ \cite{Heurtel23}). Fortunately, in the case of two qubits, Refs. \cite{PhysRevA.69.010301,PhysRevA.69.032315} guarantee that any two-qubit random operation can be produced using only three CNOT gates. Motivated by these results, we now give a protocol to generate Haar-random two-qubit unitary using a fixed-depth native gate sequence which contains $\{R_Z(\phi), R_X(\theta), CZ\}$ gates. The 15 free parameters are chosen as in Fig. \ref{fig:universal_circuits}. 
	
	%
%
	%
	The following algorithm produced Haar-random two-qubit unitaries: 
	\begin{tcolorbox}[colback=red!5!white,colframe=red!75!black]
		\textbf{Algorithm:}
		\begin{enumerate}
			\item    Generate angles $\phi_1, \phi_2, \phi_3, \phi_4$ with the probability distribution:
			\begin{align*}
				d\mu_H = \frac{1}{(2\pi)^4 4!}\Pi_{i<j} \abs{ e^{i\phi_i} - e^{i\phi_j}}^2 d\phi_1 d\phi_2 d\phi_3 d\phi_4 ,
			\end{align*}
			\item  Find $\alpha, \beta, \delta$ by the pair-wise average of any two $\{\phi\}_j$.
			\item  Tensor product two single-qubit Haar-random unitary $A \otimes B$.
			\item   Tensor product two single-qubit Haar-random unitary $C \otimes D$.
			\item  Introduce $U = (C \otimes D) V (A \otimes B)$ as the two-qubit Haar-random circuit.
		\end{enumerate}
	\end{tcolorbox}
	
	We first determine the conditional probability distributions of the 15 parameters in the decomposed circuit as shown in Fig.\ \ref{fig:universal_circuits} and then verify that the construction indeed produces random unitary matrix from the Haar distribution.

	\subsubsection{Magic basis definition}
\begin{figure*}[t]
		\includegraphics[width = \textwidth]{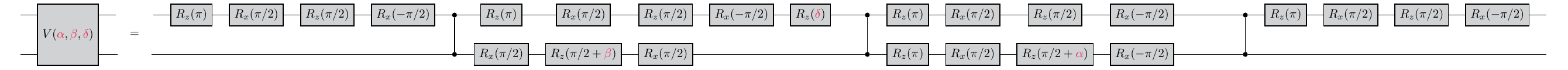}
		\caption{The construction of the entangling part $V(\alpha,\beta,\delta)$ for generating any two-qubits unitary operator as shown in Fig.\ \ref{fig:universal_circuits} using three CZ gates.}
		\label{fig:v_gate_circuit}
	\end{figure*}

	The decomposition algorithm starts with transforming $U$ into magic basis $u = \Lambda^\dagger U \Lambda$ for which $\Lambda$ is the magic matrix:
	\begin{align}
		\Lambda = \frac{1}{\sqrt{2}}
		\begin{pmatrix}
			1 & i & 0 & 0 \\
			0& 0& i &1\\
			0& 0& i & -1 \\
			1 & -i & 0 & 0 \\
		\end{pmatrix}
	\end{align}
	Since a change in basis does not contract or stretch the distribution, $u$ can be considered as a Haar-random matrix.
	From now onwards, when we write matrices with small letters mean it is in magic basis, like $v = \Lambda^\dagger V \Lambda$, where  V is the intermediate circuit standing for all multi-qubit operations (see Fig.\ \ref{fig:v_gate_circuit}).
	
	\subsubsection{Conditional distribution of $\alpha, \beta,\delta$}
	Three control parameters $\alpha, \beta,\delta$ depend on the eigenvalues of the matrix $uu^T$. Since $uu^T$ is unitary, it must have phase eigenvalues $\lambda_j  = e^{i\phi_j} \,$ for $\, j \, \in \{1,2,3,4\}$, and they can be found by the following equations \cite{shende2004minimal}:
	\begin{align}
		\begin{cases}
			\alpha &= \frac{\phi_1 + \phi_2}{2} \\
			\beta &= \frac{\phi_1 + \phi_3}{2} \\
			\delta &= \frac{\phi_2 + \phi_3}{2}
		\end{cases}
		\label{eq:first_three_params}
	\end{align}
	To determine the probability distribution of $\alpha$, $\beta$, and $\delta$, we first need to determine the probability distribution over $\phi_j$. According to Mezzadri \cite{Mezzadri}, the Haar measure induces a uniform distribution over eigenvalue phases. However, these phases are not conditionally independent. Therefore, it is necessary to find the joint probability distribution of all the eigenvalues. It has been shown that the joint probability distribution of eigenvalues for a $n$-dimensional Haar random unitary can be found using Weyl's formula \cite{Weyl}
	\begin{align}
		d\mu_H = \frac{1}{(2\pi)^n n!}\Pi_{i<j} \abs{ e^{i\phi_i} - e^{i\phi_j}}^2 d^n\phi\label{eq:joint_prob_dist_haar_eigen} 
	\end{align}
	Considering the later distribution for $u$ eigenvalues, the joint probability distribution of the same can be found for $uu^T$. In Sec.\ \ref{append:unitary_symmetric_eigenvals} we show how they come from the Haar measure.
	Sampling $\{\phi_j\}^{n=4}_{j=1}$ according to Eq. (\ref{eq:joint_prob_dist_haar_eigen}) for $n=4$
	\begin{align}
		d\mu_H = \frac{1}{(2\pi)^4 4!}\Pi_{i<j} \abs{ e^{i\phi_i} - e^{i\phi_j}}^2 d\phi_1 d\phi_2 d\phi_3 d\phi_4
	\end{align}
	and then using the Eq. (\ref{eq:first_three_params}), we get the parameter set $\{\alpha,\beta,\delta\}$ for construction of the entangling part $V$ as shown in Fig. \ref{fig:v_gate_circuit}.
	The complete construction of the Haar-random unitary $U$ is completed after generating single-qubit rotations $A \otimes B$ and $C \otimes D$ \cite{shende2004minimal}. It is essential to select each according to a single-qubit Haar-random distribution. Since the Haar measure is a rotation-invariant distribution, each of these parts is simply a single-qubit rotation.
	
	\subsubsection{$uu^T$ eigenvalues distribution}
	\label{append:unitary_symmetric_eigenvals}
	
	According to the standard unitary decomposition \cite{horn2012matrix}, a unitary matrix U can be expressed as $U = Qe^{iT}$, where $Q$ is an orthogonal matrix and $T$ is a symmetric matrix. This means that for a complex symmetric unitary matrix like $UU^T$, we can simplify it to $UU^T = e^{2iT}$.
	We can find a matrix $S$ for any complex symmetric unitary matrix $W$ such that $W = e^{2iS}$. Moreover, we can obtain $S$ by using the formula $S = \frac{1}{2i} log(W)$. By combining any member of $O(n)$, say $K$, with $e^{2iS}$, we can create a unitary matrix $J = K e^{2iS}$, where $JJ^T = W$. The set of possible unitary matrices $J$ for a given $W$ is the same as that for $O(n)$ and is therefore invariant to $W$.
	Although the function $f(u) = uu^T$ is not one-to-one, it is a many-to-one function, where the number of pre-images for any given image is infinite. This means that there are infinitely many unitary matrices that could have produced any given complex symmetric unitary matrix. The size of the possible matrix $u$ is constant across the image of the function. 
	In summary, the many-to-one property of this function implies that if we had a Haar distribution in the domain $U(n)$, then we would have the same distribution in the image of the function.
		\begin{figure}[b]
		\centering
		\includegraphics[width=0.99\linewidth]{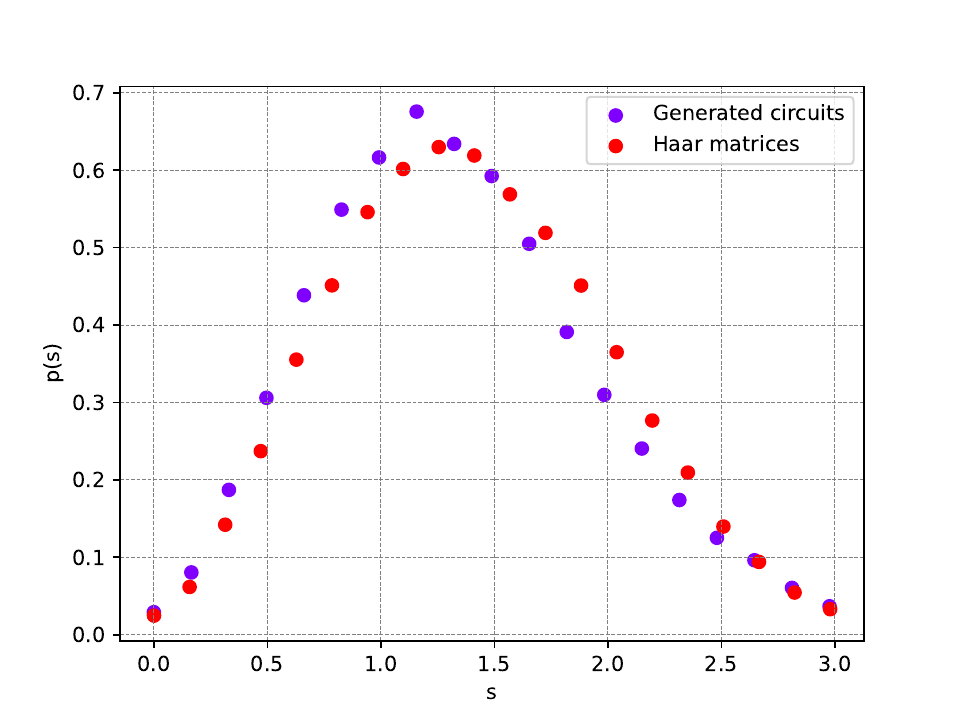}
		\caption{Probability distribution of the eigenvalue spacing of the two-qubit unitaries produced using our algorithm and that of the circular unitary ensemble (CUE).}
		\label{fig:ver2}
	\end{figure}
	
	\subsubsection{Verification}
	It is not easy to verify Haar uniformity by visualizing it on the Bloch sphere like the single-qubit case. Nevertheless, multiple verification methods exist that can confirm the resulting distribution adheres to a Haar distribution. One such test is to check the eigenvalue phase spacing of the unitary matrices generated using our protocol. The distribution of the same closely matches the circular unitary ensemble (CUE), which can confirm that the unitaries produced using our protocol are Haar-uniform. We first generate a set of two-qubit unitaries following our algorithm and calculate their eigenvalues. We plot the probabilities of the eigenvalue spacings of our generated unitaries in Fig.\ \ref{fig:ver2}. The probability distribution exhibits a close resemblance to that of the CUE.

	\begin{table*}[t]
		\centering
		\begin{tabular}{ |p{7cm}||p{2.5cm}|p{2.5cm}|p{2.5cm}|  }
			\hline
			\multicolumn{4}{|c|}{Randomized Benchmarking for Aspen-M-3} \\
			\hline
			Protocol name & Average Fidelity  & Number of different gates sequences  & Number of shots per sequence\\
			\hline
			Reported calibration data ($0^\text{th}$ qubit, f1QRB) & $99.73 \pm 0.01$ & & \\
			1Q t-design with compiler   & 0.987    &200&  800\\
			1Q with conditional probability over native gate set &   0.987  & 200   & 800 \\
			\hline
			Reported calibration data ($41^\text{th}$ and $42^\text{th}$ qubit, fCZ) & $(98.68 \pm 0.38)^3$ & & \\
			2Q t-design with compiler & 0.932 & 200 & 800\\
			2Q with conditional probability over native gate set & 0.917 & 200 & 800\\
			\hline
		\end{tabular}
		\caption{Records of different Randomized Benchmarking method. For single-qubit scenarios, the 0th qubit of the Aspen-M-3 device was used, and for two-qubit scenarios, the couple of $41^\text{th}$ and $42^\text{th}$ qubit were used due to their maximum entanglement efficiency at the time of calibration. }
		\label{tab:rb_results}
	\end{table*}

	\subsubsection{Restricted RB results}
	
	Using the fixed-depth gate sequence for each random operation, we ensure that the noise channels associated with each random operation are uniformly treated, which when compared with the compiler-optimized algorithm shows the overestimation of the latter. For the 
	{\it t design} based RB, we randomly sample from the set of two-qubit Clifford operators \cite{PhysRevA.80.012304}. To create an RB sequence of length m, we uniformly sample from the Clifford set each time which was later implemented using a fixed depth gate sequence. The inverse operations are first computed from the previous $m$ chosen Clifford element that is executed using our given algorithm. 
	We benchmark them against the calibration data. Due to the unavailability of 2-qubit randomized benchmarking data, we use an approximate way of calculating the same. We assume that the single-qubit operations are perfect and there is no associated error to its implementation. Therefore, the only noisy operations are two-qubit ones. As Refs. \cite{PhysRevA.69.010301, PhysRevA.69.032315} guarantees that only three two-qubit gates are enough to implement any arbitrary two-qubit operations, we can take two-qubit gate fidelity to the power three as an approximate estimate of average two-qubit fidelity from the calibration data. For Rigetti hardware, the average CZ gate-fidelity is given by 0.9862 (see Table \ref{tab:rb_results}). Therefore the RB estimate from the calibration data would be $0.9862^3\approx 959 (168)$. 
	
\begin{figure}[b]
		\centering
		\includegraphics[width=0.99\linewidth]{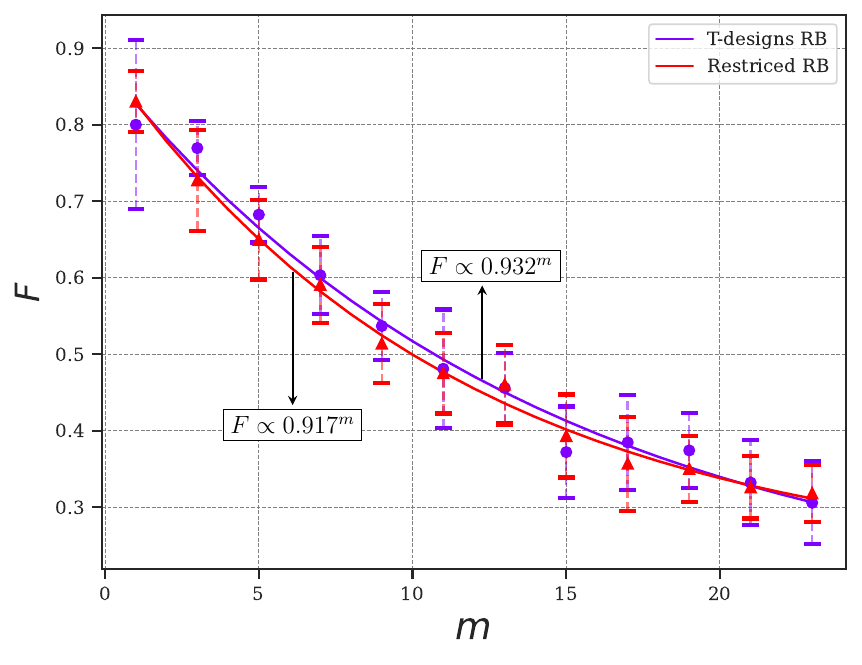}
		\captionsetup{justification=justified}
		\caption{Comparative analysis of two-qubit RB. - We present a comparison between two versions of Randomized Benchmarking (RB) - restricted RB and two-designs RB - applied to qubits 41st and 42nd of Rigetti's Aspen-M-3 quantum processor. Error bars represent the standard deviation from 200 generated circuits with 800 shots each (refer to Table \ref{tab:rb_results} for detailed results). The purple curve illustrates the extrapolated exponential fit for the two-designs RB, given by the equation $F = 0.710 \times 0.932^m + 0.166$. Similarly, the red curve represents the fit for the restricted RB, described by $F = 0.662 \times 0.917^m + 0.221$. 
		}
		\label{fig:rb_two_qubits}
	\end{figure}

	In Fig.\ \ref{fig:rb_two_qubits}, we plot the average fidelity against sequence length m for the restricted RB and the {\it t-design} RB protocol. Fitting the data to Eq. (\ref{AB}), results in the average fidelity 0.932 and 0.917 for the {\it t-design} and the restricted RB fit respectively. Comparing them with the calibration data, we can safely conclude the calibration data is clearly an overestimation of the actual two-qubit gate fidelity. It might be interesting to check-out the possibility of finding a one-to-one correspondence between the calibration data and the overestimation so that the actual performance of device can be inferred directly from the calibration data of the given native gaterset, however, that exceeds the current scopes of this study. 
	
	As mentioned earlier, the noise maps associated with the native gate implementations generally do not commute with the gate operations. Consequently, the conventional RB theory, which predicts an exponential decay of average fidelity due to the unitary invariance of the Haar-twirl, does not apply in this scenario. However, in our observed result we do observe an exponential decay of the average fidelity. While some studies \cite{Kawakami2016,Veldhorst2014,PhysRevA.92.022326,PhysRevLett.119.130502} have observed nonexponential decay potentially due to deviation from assumed noise characteristics, instances of exponential decay can still occur under different conditions~\cite{PhysRevA.89.062321, PhysRevA.103.022607, wallman2018randomized}. In Appendix \ref{append:gate_dependent_channel}, we present an analysis of our considered Haar twirl assuming that our noise channel is composed of both coherent and incoherent components which we take to be amplitude-damping and depolarizing channels. We show that the effective noise channel corrsponding to the Haar-averaged unitaries achieves gate independence when the strength of the depolarizing noise is greater than that of the depolarizing noise. We speculate that this could be one of the explanations of our observed exponential decay.
	
	An natural question about our method would the scalability of our protocol to n-qubits. We note here that generalization of our protocol beyond two qubits is not straightforward since the optimal constrction of an n-qubit unitary using any universal gateset is not known beyond two qubits. However, it is possible to ensure the fixed sequence length construction for any n-qubit unitary \cite{PhysRevLett.92.177902,Sriluckshmy2023,PhysRevResearch.5.013065}, but the optimality of the same is not guaranteed. 
	Therefore, our approach can be adapted to any number of qubits using a fixed sequence length, though it might not be an optimal construction.    
	
	\section{Summary}
	
	In this work, we introduced a new approach to randomized benchmarking by using fixed-length native gate sequences. Conventional RB methods, including the efficient {\it t-design} based ones, often overestimate the actual performance of the device. We show this rigorously and quantify the same for some exemplary scenarios.
	
	Our first contribution is the protocol to generate Haar uniform unitary for single and two-qubit operations by accomplishing a suitable choice of marginal distribution of native gate parameters. 
	%
	We have corroborated the same with different verification tests which asserts that the protocols can indeed generate fixed-depth decomposition of Haar-uniform operations for single and two-qubit scenarios. We demonstrate how Haar randomness can be practically generated, making it a resource-friendly and compiler-aware option for implementing RB on current quantum hardware. 
	Secondly, we find the average gate fidelity using two known methods and compare the same with our restricted RB technique. We show that the inference from the calibration data about the RB fidelity is not an accurate estimate of the device noise. The standard and {\it t-design} based methods also overestimate the device performance when compiler optimization is enabled. 
	This method allows for a true assessment of the hardware's performance, rather than simply trusting its calibration data or the advertised single-number metric.
	Our restricted RB protocol aims to reconcile the theoretical aspects of RB with its practical implementation, emphasizing its effectiveness and appropriateness for systems with fewer qubits.

	%
	Our observed results are robust and consistent in various trials with different sampling sequences. This implies that the method used for the sampling is reliable and does not affect the outcome of the Haar-average under different sampling conditions. This stability is crucial for validating the effectiveness of the protocols developed for generating Haar-uniform operations.
	We supplement our results with a noise analysis to elucidate on the hardware's noise profile. One plausible explanation for our observed results could be attributed to the predominant depolarizing noise within the hardware, particularly when considering a smaller number of qubits. Such an approach could be useful for building better noise models for the hardware, subsequently facilitating the development of better error mitigation strategies.

	\acknowledgements
	
	DS thanks Eliott Mamon and Marine Demarty for illuminating discussions. This work was supported in part by the Innovate UK grant Commercializing Quantum Technologies (Application No. 44167). 
	
	
	\appendix

	
	

	\section{Gate independence of the effective twirl}
	\label{append:gate_dependent_channel}
	
	In the original RB protocol, when constructing the Haar-twirl of the noise channel, we assume that each random operation $U$ is associated with a noise channel $\Lambda_U$. However, in the restricted RB protocol, each random operation is not a single operation, instead, they are made of device parametric native gates as $\{X^{(\alpha)}\}_{i,\alpha}$
	where $\alpha$ is the gate parameter and $i$ is the sequence depth. Therefore, there are associated noise channels to each of the native gate operations and it is not straightforward to see that averaging over Haar-random operation will create a unitary invariant Haar-twirled channel. 
	
	In the following, we show that even in the case of restricted RB, it is possible to construct an effective noise channel where each Haar-random operations have an effective gate-dependent noise channel which consists of the the native gates and their noise channels.

	\subsection{Effective noise channel} \label{append:effective_noise}
	\begin{figure}[b]
		\begin{subfigure}[b]{0.3\textwidth}
			\centering
			\includegraphics[width=\textwidth]{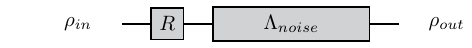}
			\caption{}
			\label{fig:gate_noise}
		\end{subfigure}
		\hfill
		\begin{subfigure}[b]{0.3\textwidth}
			\centering
			\includegraphics[width=\textwidth]{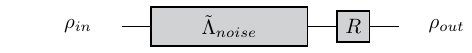}
			\caption{}
			\label{fig:reveresed_noise}
		\end{subfigure}
		\caption{Effective noise channel for a single gate $R$ following a noise channel. Here, $R$ is an arbitrary native gate, and $\Lambda$ and $\tilde{\Lambda}$ are different noise channels in a way that two ends of both circuits have the same density of quantum states.}
	\end{figure}
	\begin{figure}
		\begin{subfigure}[b]{0.5\textwidth}
			\centering
			\includegraphics[width=\textwidth]{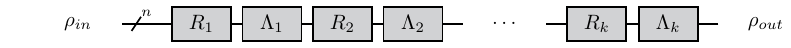}
			\caption{}
			\label{fig:real_noise}
		\end{subfigure}
		\hfill
		\begin{subfigure}[b]{0.5\textwidth}
			\centering
			\includegraphics[width=\textwidth]{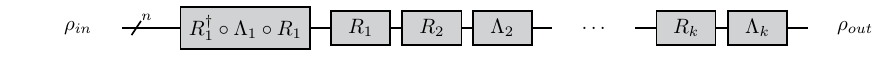}
			\caption{}
			\label{fig:real_noise_step_one}
		\end{subfigure}
		\hfill
		\begin{subfigure}[b]{0.5\textwidth}
			\centering
			\includegraphics[width=\textwidth]{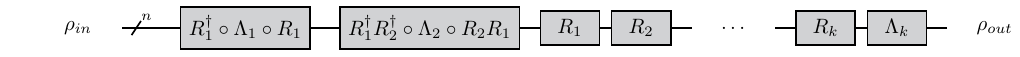}
			\caption{}
			\label{fig:real_noise_step_two}
		\end{subfigure}
		\hfill
		\begin{subfigure}[b]{0.5\textwidth}
			\centering
			\includegraphics[width=\textwidth]{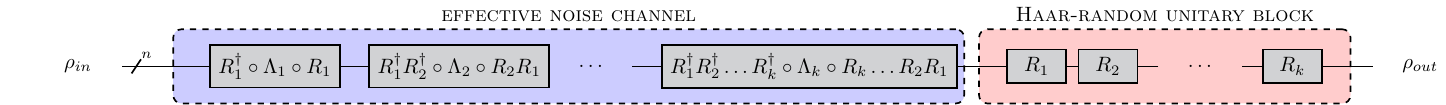}
			\caption{}
			\label{fig:effective_noise}
		\end{subfigure}
		\caption{Construction of effective noise channel for a multilayer circuit for a Haar-random operation $U$ consisting native gates $\{R_1, R_2, \ldots, R_k\}$ and their associated noise channels  $\{\Lambda_1, \Lambda_2, \ldots, \Lambda_k\}$. In practice, each gate will have its own error channel as shown in (a) which is in general non-commuting. In (b) and (c), we show a systematic iterative construction of the effective error channel to the target operation. In (d), the blue shaded region is the effective noise channel corresponding to the Haar-random operation $U$ whose native gate implementation is the green shaded region consisting of native gates.}
	\end{figure}
	

	The algorithm in the method describes a circuit composed of a sequence of native gates. Since none of them are perfect, each of native gate will undergo a distinct local noise channel. However, it can be shown that we can think of an effective noise channel for the total circuit.
	
	Let us think of a native gate $R$ implemented on an input state $\rho_{in}$. Providing  after $R$ the state goes into some noise channel (Fig.\ \ref{fig:gate_noise}), we can write the output state $\rho_{out}$ as the following:
	\begin{align}
		\rho_{out} = \sum_{l} B_l R \rho_{in} R^{\dagger} B^{\dagger}_l \label{eq:rot_noise_map}
	\end{align}
	Where $B_l$ matrices represent the corresponding Kraus operators of the noise channel. We can think of the total map as a transformation below:
	\begin{equation*}
		\rho_{in} \longrightarrow \rho_{out}    
	\end{equation*}
	\\
	Generally, $B_l$ operators do not commute with $R$. However, we can find another noise channel starting from $\rho_{in}$ and mapping to $R^{\dagger} \rho_{out} R$.
	\begin{align}
		\rho_{in} &\longrightarrow R^{\dagger}  \rho_{out} R
	\end{align}
	Since this is a completely positive map, Choi-Kraus theorem \cite{nielsen_chuang_2010} states that we can always find corresponding Kraus operators $A_k$ such that:
	
	\begin{align}
		\rho_{in} & \rightarrow R^{\dagger}\rho_{out} R \\
		R^{\dagger} \rho_{out} R &= \sum_{k} A_k \rho_{in} A^{\dagger}_k 
	\end{align}
	Where $A_k$s are the Kraus operators of the new channel (refer Fig.\ \ref{fig:reveresed_noise}). equivalently:
	\begin{align}
		\rho_{out} = \sum_{k} R A_k \rho_{in} A^{\dagger}_k R^{\dagger} \label{eq:noise_rot_map}
	\end{align}
	As long as the input and output states are positive, we can always find correct Kraus operators satisfying Eq.\ (\ref{eq:noise_rot_map}) and (\ref{eq:rot_noise_map}).  Particularly, for this single-layer circuit, $A_k$s can be found by this definition:
	\begin{align}
		A_k = R^{\dagger} B_k R
	\end{align}
	The above mentioned proof has a very practical implication. For a more generalized case in which we have a series of native gates $\{R_i\}^k_{i=1}$trying to make a Haar-random unitary circuit (Fig.\ \ref{fig:real_noise})
	\begin{align}
		&\rho_{out} \nonumber\\
		&= \sum_{i_1,i_2,...,i_k} B_{i_k} R_{k} ... B_{i_2} R_{2} B_{i_1} R_{1} \rho_{in} R^{\dagger}_{1} B^{\dagger}_{i_1} R^{\dagger}_{2} B^{\dagger}_{i_2} ... R^{\dagger}_{k} B^{\dagger}_{i_k}\\
		\intertext{with}
		&U = \Pi^{k}_{i=1} R_i
	\end{align}
	where $\Lambda_k$s represent the in-between layer noise channels. Eventually, we can continue the induction similarly till the kth-step (Fig.\ \ref{fig:effective_noise}) such that 
	\begin{align}
		&\rho_{out} =\nonumber\\ 
		&\sum_{i_1,i_2,...,i_k} A_{i_k} ... A_{i_2} A_{i_1} R_{k} ...  R_{2} R_{1} \rho_{in} R^{\dagger}_{1}  R^{\dagger}_{2} ... R^{\dagger}_{k} A^{\dagger}_{i_1} A^{\dagger}_{i_2} ... A^{\dagger}_{i_k}
	\end{align}
	Effectively, we can introduce $K_n \equiv A_{i_k} ... A_{i_2} A_{i_1}$ as the Kraus operator of an effective noise channel for the total circuit depending on circuits gates (refer Fig.\ \ref{fig:effective_noise})
	\begin{align}
		\rho_{out} = \sum_n K_n U \rho_{in} U^{\dagger} K^{\dagger}_n.
	\end{align}

	\subsection{Numerical evidence of exponential decay of average fidelity for weekly gate-dependent effective noise channel}
	
	For the standard RB, we assume that the noise associated with each Haar-random unitary is gate-independent which ultimately leads to the exponential decay of average fidelity as shown in Eq.\ (\ref{AB}). However, in our protocol of restricted RB, even if we assume that the noise channel for each native gate is gate-independent, the effective noise channel (refer Fig.\ \ref{fig:effective_noise}~) corresponding to a single Haar-random unitary is still gate-dependent as the noise channels, in general, do not commute with the gates. So, the proof for the unitary invariance of the Haar-twirled superoperator and subsequently the exponential decay of the average fidelity is not straightforward to see for restricted RB protocol. 
	
	However, as can be seen for single- and two-qubit restricted RB results (refer Fig.\ \ref{fig:rb_single_qubit} and \ref{fig:rb_two_qubits} respectively), the average fidelities do decay exponentially in our demonstration with Rigeeti ASPEN-M-3 hardware. Unfortunately, no precise error channel is known for individual qubits and native gate operations on them for the Rigetti hardware. However, assuming some noise model, it is possible to extract the noise parameters from the calibration data. The calibration data of the Rigetti ASPEN-M-3 hardware from the day of our experiments is shown in Table \ref{table:calibration}. 
	
	Following the footprints of Ref.\ \cite{PhysRevResearch.4.033140, PhysRevLett.127.270502}, we assume that the experimental noise in the hardware is mainly depolarizing ($\Lambda_d$). In addition, to model the energy decay, we assume that there is also an associated amplitude damping channel, $\Lambda_{AD}$. 
	In the following, assuming a typical noise model, we try to supplement our experimental result of exponential decay of average fidelity for our restricted RB protocol. The assumed noise model for each gate is therefore a  mixture of depolarising and amplitude-damping noise channels.  
	\bea
	\Lambda = \Lambda_d \circ \Lambda_{AD}
	\eea
	where $\Lambda_{d}$ 
	and $\Lambda_{AD}$ are described by  
	\bea
	&\Lambda_{d}&(\rho) = \lambda\rho + (1-\lambda) \Tr(\rho) \frac{\mathbb{I}}{2^n}  \label{dep}\\
	&\Lambda_{AD}& \Big( \begin{bmatrix}
		\rho_{00} & \rho_{01}\\
		\rho_{10} & \rho_{11}
	\end{bmatrix} \Big) = 
	\begin{bmatrix}
		\rho_{00} + \epsilon\rho_{11} & \sqrt{1-\epsilon}\rho_{01}\\
		\sqrt{1-\epsilon}\rho_{10} & (1-\epsilon)\rho_{11}
	\end{bmatrix}
	\label{AD}
	\eea
	respectively. 
	The parameter values of the assumed model can be identified by matching the observed average fidelities from the calibration data. For the single-qubit local depolarizing model $\Lambda_{d}$, the depolarising noise parameter turns out to be $$\lambda=2(1-F)$$ where $F$ is the {\it f1Q sim.\ RB} calibration data \cite{Marine_entropybenchmarking}. 
	As can be verified from the calibration data in Table \ref{table:calibration}, the single-qubit gate fidelities are typically one order of magnitude higher than the two-qubit CZ gate fidelity. As fidelity loss due to single-qubit gate operations is insignificant as compared to the two-qubit gate operations, we further assume that the single-qubit gates are perfect and there is no associated noise channel. The noise in the RB experiment is solely contributed by the two-qubit gate channel that contains depolarizing and amplitude-damping channels of the following form
	\bea
	\Lambda^{2Q} = (\Lambda_d^{1} \circ \Lambda_{AD}^1)\otimes(\Lambda_d^2 \circ \Lambda_{AD}^2)
	\eea

	
	
	We now show by taking this fairly simple yet realistic noise model that the effective noise channel in the case of restricted RB protocol is indeed weakly gate-dependent which is a prerequisite of the twirling and the exponential decay of the RB fidelity. 
	
	\begin{table*}[t]
		\begin{subtable}{\textwidth}
			\setlength{\tabcolsep}{5pt}
			\begin{tabular}{lrrrrrrl}
				\toprule
				Qubit & T1 ($\mu$s) & T2 ($\mu$s) & f1QRB & f1Q sim. RB & fActiveReset & fRO & Date \\
				\midrule
				0 & 10.02 & 18.16 & 99.73\% $\pm$ 0.01\% & 99.56\% $\pm$ 0.02\% & 97.5\% & 91.9\% & 21st July 2023 \\
				\bottomrule
			\end{tabular}
		\end{subtable}
		\vspace{1em} 
		\begin{subtable}{\textwidth}
			\caption{Two-qubits device calibration}
			\label{tab:two_qubits_calibration}
			\setlength{\tabcolsep}{5pt}
			\begin{tabular}{lrrrrrrrrl}
				\toprule
				Pair & fXY & fXY std err & fCZ & fCZ std err & Avg T1 (µs) & Avg T2 (µs) & Avg fActiveReset & Avg fRO & Date \\
				\midrule
				41-42 & 0.9688 & 0.004721 & 0.9868 & 0.003863 & 32.4 & 50.6 & 0.994 & 0.9105 & 19th July 2023 \\
				\bottomrule
			\end{tabular}
		\end{subtable}
		\caption{Calibration data of the considered qubits of Rigetti ASPEN-M-3 hardware on the day of the experiment.}
		\label{table:calibration}
	\end{table*}
	
	In Ref.\ \cite{wallman2018randomized}, it was shown that in the case of gate-dependent noise, one can expect an exponential decay fidelity in RB provided that the gate dependency of the associated noise channel is weak. We here provide a quantification of this weakness by the diamond norm distance between two noise channels. We provide numerical evidence to show that when the value of noise parameters is small, the effective noise channel in restricted RB is independent.  
	
	\begin{figure}[h]
		\includegraphics[width=1.1\linewidth]{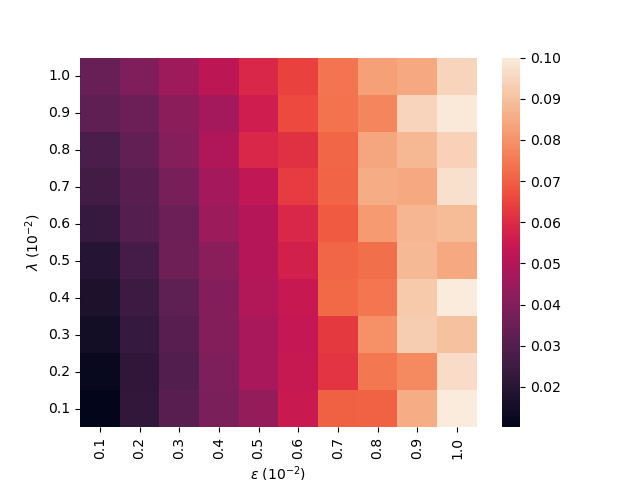}
		\captionsetup{justification=justified}
		\caption{ The average diamond norm distance between pairs of effective noise channels, corresponding to two independent Haar-random unitaries, was calculated. This average was derived from 10 pairs of independent Haar-random unitaries, a number sufficient to reduce the standard error for each cell by one order of magnitude. 
			The ordinate and abscissa are respectively the depolarising and amplitude damping noise strength as defined in Eq.\ (\ref{dep}-\ref{AD}). The color-coded value of the average diamond norm distance is presented in the legend on the right. The effective noise channel is gate independent when the diamond norm distance is zero which is the case when $\epsilon=\lambda=0$ i.e.\ in the absence of any noise. With increasing the value of the amplitude damping parameter $\epsilon$, the effective noise channel started to become gate-dependent. However, the weak gate dependency appears to be much more robust against depolarising noise strength $\lambda$.}
		\label{fig:diamond}
	\end{figure}
	
	The diamond norm distance is a well-established measure of channel discrimination \cite{Paulsen2003}. If $\Lambda_a$ and $\Lambda_b$ are two channels acting on two-qubit operations, the diamond norm distance is defined as 
	\bea
	&d_{\diamond}(\Lambda_a,\Lambda_b)& = ||\Lambda_a -\Lambda_b||_\diamond \nonumber \\
	&=& \max_\rho || (\mathbb{I}_{4}\otimes \Lambda_a)\rho -(\mathbb{I}_{4}\otimes \Lambda_b)\rho||_1
	\eea
	where the maximization is done over all density matrix $\rho$ of dimension $2^4$. In restricted RB, $\Lambda_a$ and $\Lambda_b$ are two effective noise channels corresponding to two independent Haar-random unitaries. Note that the effective noise channels are explicitly dependent on the native gates which constitute the Haar-random unitary. We will now show that even though the individual effective the error channels are gate-dependent, their averages over the Haar measure can be weakly gate-dependent under certain conditions. In Fig.\ \ref{fig:diamond}, we plot the average diamond distance $\bar{d_\diamond}$ against the depolarising and amplitude-damping noise strength.  The average diamond norm distance $\bar{d_\diamond}$ can be written as
	\begin{align}
		\bar{d_\diamond} = \frac{1}{n_p}\sum_i d_\diamond (\Lambda_{a_i}, \Lambda_{b_i}) ,
	\end{align}
	where $(\Lambda_{a_i}, \Lambda_{b_i})$ represent pairs of random effective noise channels generated according to Haar measure, and $n_p$ denotes the total number of sampling pairs. The average is taken by creating many independent Haar-random unitaries and their respective effective error channel.
	\\
	
	As can be seen from Fig.\ \ref{fig:diamond}, the effective noise channel is becoming increasingly gate-dependent as we increase the amplitude damping noise parameter $\epsilon$. However, the depolarising noise is comparatively less fatal. Even with a moderate value of depolarising noise parameter $\lambda$, the effective noise channel remains weakly gate-dependent. Therefore, as long as the principal source of our experimental noise is depolarising, we should expect an exponential decay of average fidelity 
	as observed in Fig.\ \ref{fig:rb_single_qubit} and \ref{fig:rb_two_qubits}. 
	The analytical proof for gate dependent noise twirling \cite{wallman2018randomized} holds in this case as well since the Haar measure is a 2-design. Following a similar treatment, one can show that the Haar-twirl in this case becomes a linear map composed of a depolarizing channel and a channel that reduces the trace. Therefore, even when the noise superoperator is both gate-dependent and trace-preserving, the average fidelity, when averaged over the Haar measure, corresponds to an exponential decay with an additional single perturbation term.

	\section{Generators of Clifford members} \label{append:clifford}
	
	Three fundamental gates, namely the Hadamard, the phase gate S, and the CNOT, generate the Clifford group. Their matrix form in the computational basis is given by
	\begin{align}
		&H = 
		\frac{1}{\sqrt{2}} \begin{bmatrix}
			1 & 1\\
			-1 & 1
		\end{bmatrix},  \hspace{1cm}
		S = 
		\begin{bmatrix}
			1 & 0\\
			0 & e^{i \pi/2}
		\end{bmatrix}, \hspace{1cm}\\
		&CNOT = 
		\begin{bmatrix}
			1 & 0 & 0 & 0 \\
			0 & 1 & 0 & 0 \\
			0 & 0 & 0 & 1 \\
			0 & 0 & 1 & 0 \\
		\end{bmatrix}. \label{eq:clifford_member}
	\end{align}
	\\
	\\
	
	%

	\bibliography{RB.bib}
	
\end{document}